\documentclass[11pt]{article}
\usepackage[a4paper, top=2.0cm, bottom=2.0cm, left=2.0cm, right=2.0cm]{geometry}
\usepackage{multicol}
\usepackage{setspace}
\usepackage[scaled]{helvet}
 
\usepackage[T1]{fontenc}
\usepackage{graphicx}
\usepackage{amssymb}
\usepackage{epstopdf}
\usepackage{wrapfig}
\usepackage{color}
\usepackage{chemformula}
\usepackage{caption}
\usepackage{upgreek}
\usepackage[dvipsnames]{xcolor}
\usepackage[font={color=black},figurename=Fig.,labelfont={it}]{caption}
\usepackage{siunitx}
\usepackage{authblk}
\usepackage{hyperref}
\usepackage{xcolor}
\usepackage{enumitem} 
\usepackage{paralist}
\setitemize{noitemsep,topsep=0pt,parsep=0pt,partopsep=0pt,leftmargin=1.5\parindent}
\hypersetup{
colorlinks=true,
linkcolor=blue,
urlcolor=blue,
citecolor=blue,
}
\usepackage{natbib}
\usepackage{aasmacros}
\usepackage{doi}
\usepackage[labelformat=empty]{caption}

\usepackage{soul}

\newcommand{\squishlist}{
 \begin{list}{$\bullet$}
  { \setlength{\itemsep}{1pt}
     \setlength{\parsep}{0pt}
     \setlength{\topsep}{1pt}
     \setlength{\partopsep}{0pt}
     \setlength{\leftmargin}{1.5em}
     \setlength{\labelwidth}{1.5em}
     \setlength{\labelsep}{0.5em} } }
\newcommand{\squishend}{
  \end{list}  }

\begin{document} 
\title{Exoplanet characterization with NASA's Habitable Worlds Observatory \textit{(thematic area: Astro)}}
\author[1]{\textbf{Joanna K. Barstow}*}
\author[2]{Beth Biller}
\author[3]{Mei Ting Mak}
\author[2]{Sarah Rugheimer} 
\author[4]{Amaury Triaud}
\author[5]{Hannah R. Wakeford}
\affil[1]{School of Physical Sciences, The Open University, Walton Hall, Milton Keynes, MK7 6AA}
\affil[2]{Institute for Astronomy, University of Edinburgh, Royal Observatory, Edinburgh, EH9 3HJ}
\affil[3]{Atmospheric, Oceanic and Planetary Physics, Department of Physics, University of Oxford, Clarendon Laboratory, Parks Road, Oxford, OX1 3PU}
\affil[4]{School of Physics \& Astronomy, University of Birmingham, Edgbaston, Birmingham, B15 2TT}
\affil[5]{School of Physics, HH Wills Laboratory, Tyndall Avenue, Bristol, BS8 1TL}
\affil[*]{Contact: Jo.Barstow@open.ac.uk}
\date{}

\maketitle

\section{Scientific Motivation \& Objectives}
% \begin{itemize}
%     \item Transformative exoplanet science is to characterise true solar system analogue planets. 
%     \item Studying these objects may lead to the discovery of other habitable worlds (or inhabited) but also allows us to answer questions about the rarity of Earths.
%     \item Studying other, similar planetary systems places the solar system in context and informs about planet formation.
%     \item HWO is the mission that provides a pathway for both habitable worlds characterization and also solar system equivalents. It is a UV-optical-IR telescope, this makes it feasible to study rocky and gas giant planets in reflected light.
% \end{itemize}

Exoplanet atmosphere characterization has seen revolutionary advances over the last few years, particularly with the launch of JWST. With its sensitivity and infrared wavelength coverage, JWST has provided crucial spectroscopic observations of numerous transiting exoplanets that orbit close to their parent stars \citep[e.g.][]{jwst_tec2023,grant2023,Madhusudhan_2023_K218b,evans-soma2025}, as well as directly imaging young, self-luminous planets at wide separations \citep[e.g ][]{carter2023,balmer2025}. Ground-based instruments such as the Very Large Telescope GRAVITY instrument also continue to provide detailed near-infrared spectra of young gas giants \citep[e.g.][]{Nasedkin2024}, and high resolution spectrographs such as ESPRESSO allow us to probe global chemistry and winds on some of the most extreme hot Jupiters \citep[e.g.][]{seidel2023,prinoth2025}. Whilst these observations and subsequent analysis efforts are providing us with unique insights into atmospheric chemistry, dynamics and planet formation mechanisms, with current instrumentation true solar system analog planets remain inaccessible. A major goal for exoplanet science over the coming decades is to observe, and characterize, temperate rocky planets and cool gas giants in orbit around solar-type stars. 

A key motivator for this goal is of course the prospect of detecting signs of habitability, or perhaps even signs of life, on another rocky planet. Such a discovery would profoundly impact our understanding of our place in the universe, and allow us to begin answering the question of how common (or rare) Earth-like (or indeed inhabited) worlds are. A full understanding of the origins of our solar system also requires us to study analogs of planets such as Venus, Jupiter and Saturn.  

Characterization and categorization of these planets relies on spectroscopic observations capable of identifying molecular species in their atmospheres. Such observations may be conducted in the ultraviolet and optical (measuring reflected light from the parent star) or the infrared (measuring thermal emission from the planet itself). Regardless of the wavelength range, these observations represent a substantial engineering challenge; the contrast between a temperate, Earth-sized exoplanet and its parent star is on the order of 10$^{-10}$ in the optical, and even for larger gas giants the contrast is still quite extreme (10$^{-9}$ -- 10$^{-8}$). This requires sophisticated telescope optics that can effectively block the signal from the central star and allow the planet to be directly observed.

NASA's next flagship mission, the Habitable Worlds Observatory (HWO), is designed to overcome this difficulty. HWO is planned for launch in the mid-2040s. Whilst HWO is a general-purpose observatory, it will boast a coronagraphic instrument capable of reaching the needed 10$^{-10}$ contrast, on an ultrastable platform enabling long integration times to achieve the required signal to noise. HWO will cover wavelengths from the near-ultraviolet to the near-infrared, enabling detections of key biosignature molecules such as O$_2$, O$_3$ and CH$_4$, and also allowing study of clouds and detection of key habitability indicators such as ocean glint \citep{vaughan2023} and a vegetation `red edge' \citep{omalley_james2018}. HWO will also enable detection of sulphur species in the UV (critical for studying Venus analogs) and allow us to determine the abundance of chemicals such as CH$_4$, NH$_3$ and PH$_3$  on cold gas giants resembling Jupiter and Saturn. The observatory's wavelength range is set at the reddest end by thermal noise considerations, since unlike infrared telescope JWST the observatory will be at room temperature.

Via early involvement in both science and engineering aspects of this groundbreaking observatory, the UK exoplanet community now has an important opportunity to influence the telescope's design and ensure that it enables us to achieve our science goals. There is also an excellent prospect for UK experts in instrument and detector design and manufacture to contribute hardware to the mission, an endeavour that we support.
To maintain our international competitiveness, the UK exoplanet community must be at the forefront of observational campaigns with HWO when it eventually launches, and this comes with the need for parallel development in laboratory astrophysics and computational modelling to ensure readiness. Maximising our exploitation of this transformative NASA mission requires consistent financial support in all of these areas across the next two decades.

\section{Strategic Context}
% \begin{itemize}
%     \item HWO will be NASA's next flagship - v. important for UK scientists and engineers to be embedded early
%     \item Early involvement creates opportunity to influence direction of travel and also to get benefits from providing hardware (UK companies winning contracts)
% \end{itemize}

HWO is NASA's next flagship observatory, following the legacy of Hubble and JWST. UK astronomers and in particular the exoplanet community have been very successful in exploiting both of these missions, and there was substantial UK hardware involvement in JWST, with the UK Astronomy Technology Centre leading the design and manufacture of its Mid InfraRed Instrument, MIRI \citep{wright2023} 

A key point of interest for exoplanet scientists is that HWO is the first NASA flagship observatory that will be developed with exoplanet characterization as a key goal. Whilst both Hubble and JWST have allowed us to make immense strides in this area, neither telescope is specifically optimized for exoplanet observations, since Hubble was already in flight and JWST was advanced in development before exoplanet transit spectroscopy and coronagraphic imaging gained traction in the 2010s. Whilst the upcoming Nancy Grace Roman Space Telescope \citep{bailey2023}will significantly advance exoplanet demographic studies via detection, characterization of specific targets with the coronagraph forms only a small fraction of the overall mission, with a total of just 90 days allocated to this instrument in the primary mission. The opportunity for exoplanet characterization science goals to drive instrument development for this observatory is therefore unprecedented, and we must take full advantage. 

During the Voyage 2050 prioritization exercise, the European Space Agency identified observaion of temperate exoplanets as a key science driver for a future L-class mission (see the \href{https://www.cosmos.esa.int/documents/1866264/1866292/Voyage2050-Senior-Committee-report-public.pdf}{Voyage 2050 report}). An emerging mission concept in response to this is the Large Interferometer for Exoplanets \citep[LIFE,][]{Quanz2022_LIFEI}, which will target biosignatures at mid-infrared wavelengths. HWO and LIFE are highly complementary since they probe very different wavelength regions, sensitive to different aspects of temperate atmospheres, that could not possibly be accessed from a single telescope due to these wavelength regions demanding different engineering solutions. In addition, the strategic importance of HWO as a general-purpose UV-optical-IR observaotry cannot be overstated. Currently our access to UV wavelengths in particular is provided by Hubble, but with more than 35 years of operational history, Hubble's lifetime is now limited, with deorbit likely in the mid-2030s. HWO is required to restore our access to UV skies following Hubble's mission end.

Although HWO is not due to launch until the 2040s, a key date is the Mission Concept Review (MCR), due to take place at the end of 2029. By the MCR, the aim is for all critical technologies to be at a technology readiness level (TRL) of at least 5. By then, first-generation instruments will be identified and the overall telescope architecture will be set. Critical science drivers for instrument and telescope parameters must therefore be identified and communicated within the next few years. 

\section{Proposed Approach}
% \begin{itemize}
%     \item Direct imaging with HWO coronagraph
%     \item transiting exoplanet science with HWO
%     \item Exoring observations
%     \item biosignature identification/checks for false positives
%     \item spectroscopy, phase curves/multi-phase observations, polarimetry
% \end{itemize}

\begin{figure}   \includegraphics[width=1.0\textwidth]{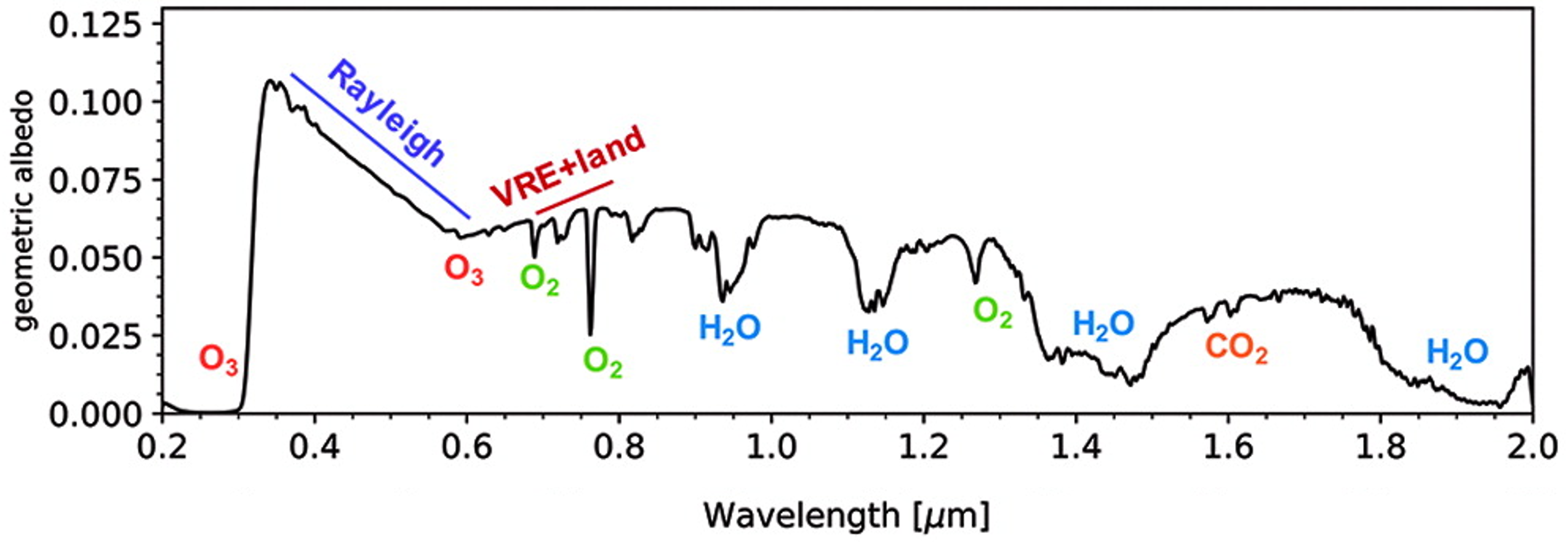} \caption{Figure 1 is taken from \cite{schwieterman} "Exoplanet Biosignatures: A Review of Remotely Detectable Signs of Life", Figure 4, published in Astrobiology and reproduced with permission. A synthetic UVO-optical Earth radiance spectrum at quadrature phase (half illumination) in terms of geometric albedo. This spectrum was generated by the VPL 3D spectral Earth model \citep{robinson2011,schwieterman2015}. Strong absorption features from O$_2$, O$_3$, H$_2$O, CO$_2$, N$_2$O, and CH$_4$ are labeled, in addition to Rayleigh scattering and the location of the vegetation red edge (VRE).} \label{earth_spectrum}
\end{figure}

\begin{figure}
\centering
    \includegraphics[width=0.8\textwidth]{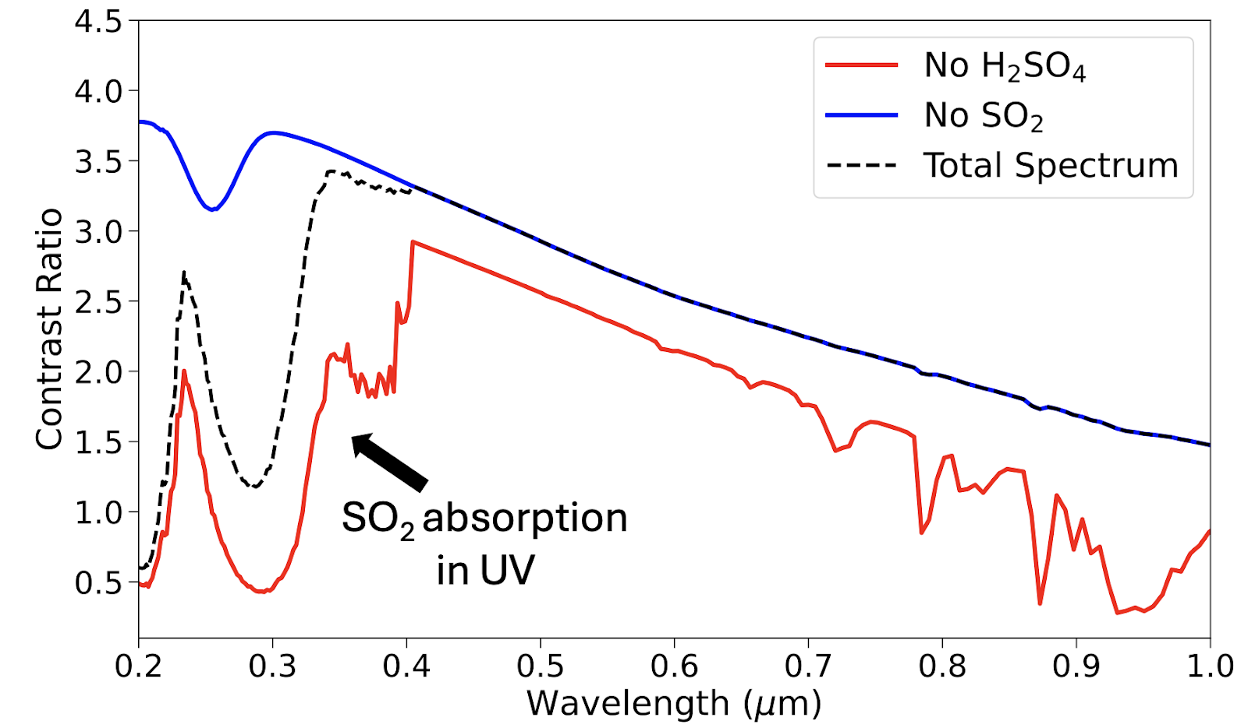} \caption{Figure 2 is taken from the Science Case Development Document \href{https://docs.google.com/document/d/18rb6H7SNd15Is_NzoQkmoaqEocDkcVt5/edit\#heading=h.rxylahxbnjx8}{Detecting and Characterizing Venus Analogs Orbiting Other Stars}. This is a simulated reflected light spectrum in the optical and ultraviolet showing the observable effects of SO$_2$ and H$_2$SO$_4$ clouds. The figure is reproduced here with the permission of S. R. Kane.} \label{venus_spectrum}
\end{figure}

Exoplanet characterization with HWO will combine many different observational strategies, which are described in more detail within \href{https://docs.google.com/spreadsheets/d/1PTazkPP-gIhOEETNVDLoXp-7m-1etTRdknWWlmQrPNI/edit?gid=1800635110#gid=1800635110}{Science Case Development Documents} written by the HWO START community. Observations of both rocky and gas giant planets in reflected light will be carried out using the coronagraphic instrument that is central to the mission; this instrument will enable imaging and spectroscopy from near-ultraviolet to near-infrared wavelengths, with the details of wavelength cutoffs and resolving power still to be defined. Simulated observations of Earth (Figure~\ref{earth_spectrum}) and Venus (Figure~\ref{venus_spectrum}) are shown, demonstrating what may be learned about rocky planets from coronagraphic observations in reflected light, with spectra of Earth-like exoplanets probing O$_3$, O$_2$, H$_2$O and and Venus-like spectra being sensitive to sulphur gases and aerosols. Such observations will allow us to classify and discriminate between different types of rocky atmospheres (see SCDD \href{https://docs.google.com/document/d/1bQFVPS1PacCraszTA_fI5CXhk0tAQujm/edit#heading=h.gjdgxs}{The Search for Life on Potentially Habitable Exoplanets},\citealt{triaud2024}. Similar observations of cold gas giants will allow us to investigate cloud formation processes by measuring the depth of CH$_4$ absorption features across the HWO wavelength range and investigating ultraviolet absorption from sulphur, phosphorus and nitrogen species (see SCDD \href{https://docs.google.com/document/d/1dFH3SK1OOi1UMWS-Bh8PrJljK-hojaRk/edit#heading=h.3znysh7}{Direct Imaging Characterization of Cool Gaseous Planets}). 

Transiting exoplanets and those in close to edge-on orbits will not be neglected. Whilst these targets will not be as well suited to coronagraphic observations since they will mostly be inside the inner working angle of the instrument, an imager with spectroscopic capability is identified as another instrument for the telescope (see Section~\ref{tech}) and will be suitable for observations of transiting planets. With its highly stable platform, HWO is well suited to obtaining phase curve measurements that track the planet throughout an orbit, monitoring the changes in overall system flux as the illuminated side of the planet rotates in and out of view. If a broad enough wavelength range is covered, this allows direct constraints on the planet's energy budget and heat redistribution. Currently restricted to performing these measurements only for ultra-hot planets with very short periods, with HWO we hope to extend these observations to periods of several tens of days, enabling us to probe the changing atmospheric dynamics of gas giants as distance from the parent star increases \citep{wakeford2025}. The phase curve technique will also be employed for directly imaged planets via the coronagraph; these planets will not be tidally locked, and measuring their brightness as a function of time over a period of several days can provide constraints on their rotation period, recover signatures of atmospheric variability \citep[e.g.][]{vos2022}, and indicate the presence of different types of surface on rocky planets \citep[e.g.][]{robinson2025}.

Whilst polarimetric capability is not a guarantee for HWO, many science cases rely on, or would be enhanced by, polarimetric measurements. This includes science cases for planets directly imaged by the coronagraph, and also transiting planets that may be accessed via phase curves with a spectroscopic imager or dedicated spectropolarimeter. The wavelength-dependent degree of linear polarization is affected by the presence of scattering particles in the atmosphere, making it a useful discriminator for different cloud compositions \citep[e.g.][]{chubb2024}. Polarimetric signatures can also help to reveal the presence of surface liquid water \citep{cowan2025}. 
Several exoplanet science cases also require observations at high spectral resolving power, particularly in the UV. The UV of the host star strongly determines the photochemistry dominant on an exoplanet atmosphere \citep[see e.g.][]{segura2003, rugheimer2013} and after HST retires, HWO will be our next UV mission to get these critical observations to interpret our atmospheric results. Such measurements allow us to probe exoplanet exospheres and atmospheric escape by measuring the strength of narrow atomic lines (e.g. SCDD \href{https://docs.google.com/document/d/1j758X9m8a-84CpiQWegAH9wO_kTPaI_Lx35QHrRrKd4/edit?tab=t.0#heading=h.2et92p0}{Atmospheric Escape and Habitability}). High spectral resolving power (R$\sim$ 100,000) can also enable the measurement of wind speeds from Doppler shifts of molecular bands, placing strong constraint on atmospheric dynamics and transport of materials.

As well as observations of planets themselves, HWO will also be able to observe exoring systems by measuring excess brightness for a giant planet in reflected light (SCDD \href{https://docs.google.com/document/d/1Fs07W1XQvR4uy0nm5jkG4Npf8hB_PbrVdO8MuJkYo94/edit?tab=t.0#heading=h.ek6wyshtwzwp}{Exoring Detection, Characterization and Occurrence Rates}) and potentially moons of giant exoplanets (SCDD \href{https://docs.google.com/document/d/1bOa97ZAblcfkfIOmzQDZDYDxJaMN-Cue8MFVjDEDTqw/edit?tab=t.0#heading=h.ek6wyshtwzwp}{Exomoon Detection with Mutual Events}). 

\section{Proposed Technical Solution and Required Development}
\label{tech}
% \begin{itemize}
%     \item Support for HWO
%     \item Specific instruments: need coronagraph, expand IR capability? UV instruments - imager/MOS, UK may build one of these. 
% \end{itemize}

UK support for both technical and scientific involvement in HWO is critical for the success of UK exoplanet research in the coming decades. Especially during this current crucial period of mission development in the run up to the MCR, investigation into science drivers and technological innovations must proceed in parallel, and be funded accordingly. Support for a UK instrument contribution to HWO will foster stronger avenues of influence for scientists during development, guarantee a share of mission time for the instrument team, and also bring economic benefits to UK companies involved in manufacture. The coronagraphic instrument will be led by NASA, with the potential for hardware contributions from other agencies; there are also slots for potentially three further instruments, two of which are likely to be a high resolution imager with some spectroscopic capability and a multi-object spectrograph. Instrument studies for these two options are already funded and being undertaken by two separate UK teams. Whilst many exoplanet science goals will be most usefully met with the coronagraphic instrument, high resolution spectroscopy and transit/phase curve measurements will likely require one of these two instruments. 

In addition to development for HWO itself, preparation for the eventual interpretation of data from HWO is equally critical. This relies on complementary observations, laboratory work and computational development. Requirements for adequate model representations of exoplanet atmospheres include, but are not limited to: accurate gas absorption data, typically calculated from quantum mechanical first principles \citep[see the work of the ExoMol project and others,][]{tennyson2024} and benchmarked against laboratory measurements; suitable representations of cloud, including composition and particle size, which rely heavily on laboratory studies \citep[e.g.][]{he2020}; appropriate chemical networks, which require computational simulations informed by laboratory observations \citep[e.g.][]{fisher2025}; suitable parametric (retrieval) models with the appropriate prior representations \citep[e.g.][]{damiano2021,damiano2022}; more complex global climate simulations  \citep[e.g. ][]{kopparapu2019}; and models that capture the interplay between atmosphere, interior and biosphere \citep[e.g. ][]{triaud2024}. Addressing these points requires long-term, strategic financial support for laboratory astrophysics projects in the above areas, and also for high performance computing provision. In addition, ground-based observing campaigns are also important, for example in providing constraints on planet masses in HWO candidate systems to rule out giant planets in the habitable zone \citep{wittenmyer}; synergy between ground- and space-based astronomy must not be neglected. 

\section{UK Leadership and Capability}
% \begin{itemize}
%     \item UK already heavily involved in mission, including several exoplanet scientists having served in START etc. 
%     \item Multiple scientists on teams funded for instrument studies
%     \item Huge exoplanet atmospheres community in UK that will be served by involvement in HWO
% \end{itemize}

The UK exoplanet community has expanded in recent years, providing us with the person power to be internationally competitive. Exoplanet research groups with at least one permanent member of academic staff are present at: Cardiff University, Imperial College London, Keele University, Kings College London, Queen Mary University of London, Queens University Belfast, The Open University, University College London, University of Birmingham, University of Bristol, University of Cambridge, , University of Central Lancashire, University of Edinburgh, University of Exeter, University of Leeds, University of Leicester, University of Hertfordshire, University of Manchester, University of Oxford, University of St Andrews, and University of Warwick. Collectively, these groups cover all areas of exoplanet research, from observational detection and characterization to atmospheric and population modelling. This huge community will be well-served by investment in HWO and associated activities. 

Members of the UK exoplanet community are already embedded in the HWO development effort, with some having served on START working groups and several others involved in the ongoing UKSA-sponsored instrument studies. In terms of instrument development for a NASA mission, the UK is very well placed here, with the precedent of having contributed the MIRI instrument for JWST. As well as the vast collective experience of the exoplanet community, the UK also boasts expertise in several key technologies required for HWO, including ultraviolet detector technology and optics for imaging and spectroscopy. Contributing an instrument and/or leading a significant hardware contribution for HWO will draw on the synergy between UK exoplanet and space science communities.

\section{Partnership Opportunities}
% \begin{itemize}
%     \item Obviously partnership with NASA
%     \item Bilateral instrument development
%     \item Potential opportunities for involvement in SAGs as part of the HWO SIG. 
% \end{itemize}

Involvement in the HWO mission will of course involve partnership with NASA, but there are also other opportunities to explore. There is the possibility that any instrument contribution could be bilateral, with the UK working alongside another individual country, or leading a contribution with inputs from a number of European Space Agency partners as was the case for MIRI on JWST. 

On the science side, membership of the HWO Science Interest Group is open to all and several UK scientists are already members. Involvement in Science Analysis Groups is also possible; these are groups coordinated over a finite period of time, usually approximately two years, to address a particular science question, and several are already in progress specifically in preparation for HWO. 

%\section{Suggested Mission Class (optional)}

\section*{Co-signatories}
 \textbf{University of Arkansas:}		
Amirnezam	Amiri;
\textbf{Cardiff University:}		
Lorenzo V. 	Mugnai,
Subhajit	Sarkar;
\textbf{Cornell University/University of Bristol:}		
Lili	Alderson;
\textbf{Imperial College London:}		
Subhanjoy	Mohanty;
\textbf{Indian Institute of Astrophysics:}		
Preethi	Karpoor;
\textbf{MSSL/UCL:}		
Vincent	Van Eylen;
\textbf{Queen Mary University of London:}	
Matthew	Batley,
Edward	Gillen;
\textbf{STFC RAL Space:}		
Andrzej	Fludra;
\textbf{The Open University:}		
David	Arnot,
Carole	Haswell,
Jesper	Skottfelt,
Julia	Semprich,
Eleni	Tsiakaliari;
\textbf{University College London:}
Jonathan	Tennyson,
Sergei N. Yurchenko;
\textbf{University of Bern:}		
Marrick	Braam,
Joost P.	Wardenier;
\textbf{University of Birmingham:}		
Annelies	Mortier,
Vatsal	Panwar,
Anjali	Piette,
Adam	Stevenson;
\textbf{University of Bristol:}		
Katy L.	Chubb,
Charlotte	Fairman,
Daniel	Valentine;
\textbf{University of Cambridge:}		
Edouard	Barrier,
Amy	Bonsor,
Simon	Hodgkin,
Nikku	Madhusudhan,
Lalitha	Sairam,
Samantha J. Thompson;
University of Chicago:		
Dominic	Samra;
\textbf{University of Edinburgh:}		
Mariangela	Bonavita,
Xueqing	Chen,
Trent	Dupuy,
Aiza	Kenzhebekova,
Adam	Koval,
David A. 	Lewis,
Larissa	Palethorpe,
Paul	Palmer,
Mia Belle	Parkinson,
Ken	Rice,
Colin	Snodgrass,
Ben	Sutlieff;
\textbf{University of Exeter:}		
Sasha	Hinkley;
\textbf{University of Glasgow:}		
Matthew I.	Swayne;
\textbf{University of Leeds:}	
Richard	Booth;
\textbf{University of Leicester:}	
Martin	Barstow,
Matthew	Burleigh,
Sarah L. Casewell;
\textbf{University of Oxford:}	
Suzanne	Aigrain,
Jayne	Birkby,
Claire	Guimond,
Namrah	Habib,
Thaddeus D.	Komacek;
\textbf{University of St Andrews:}		
Andrew Collier	Cameron,
Ryan	MacDonald;
\textbf{University of Warwick:}	
David J. A. Brown,
Heather	Cegla,
Alastair Claringbold

\end{document}